\documentstyle[twoside,fleqn,espcrc2]{article}
\input prepictex
\input pictex
\input postpictex

\newdimen\xposition
\newdimen\yposition
\newdimen\dyposition
\newdimen\crossbarlength

\def\puterrorbar at #1 #2 with fuzz #3 {%
  \xposition=\Xdistance{#1}
  \yposition=\Ydistance{#2}
  \dyposition=\Ydistance{#3}

\setdimensionmode
\put {$\bullet$} at {\xposition} {\yposition}

\dimen0 = \yposition
  \advance \dimen0 by -\dyposition
\dimen2 = \yposition
  \advance \dimen2  by \dyposition
\putrule from {\xposition} {\dimen0}
  to {\xposition} {\dimen2}

\dimen4 = \xposition
  \advance \dimen4 by -.5\crossbarlength
\dimen6 = \xposition
  \advance \dimen6 by  .5\crossbarlength
\putrule from {\dimen4} {\dimen0} to {\dimen6} {\dimen0}
\putrule from {\dimen4} {\dimen2} to {\dimen6} {\dimen2}
\setcoordinatemode}

\newdimen\xpositionc
\newdimen\ypositionc
\newdimen\dypositionc
\newdimen\crossbarlength

\def\putcirclebar at #1 #2 with fuzz #3 {%
  \xpositionc=\Xdistance{#1}
  \ypositionc=\Ydistance{#2}
  \dypositionc=\Ydistance{#3}

\setdimensionmode
\put {$\circ$} at {\xpositionc} {\ypositionc}

\dimen0 = \ypositionc
  \advance \dimen0 by -\dypositionc
\dimen2 = \ypositionc
  \advance \dimen2  by \dypositionc
\putrule from {\xpositionc} {\dimen0}
  to {\xpositionc} {\dimen2}

\dimen4 = \xpositionc
  \advance \dimen4 by -.5\crossbarlength
\dimen6 = \xpositionc
  \advance \dimen6 by  .5\crossbarlength
\putrule from {\dimen4} {\dimen0} to {\dimen6} {\dimen0}
\putrule from {\dimen4} {\dimen2} to {\dimen6} {\dimen2}
\setcoordinatemode}

\newcommand{\AmS}{{\protect\the\textfont2
  A\kern-.1667em\lower.5ex\hbox{M}\kern-.125emS}}

\title{Gauge Invariance and Confinement in Noncompact Simulations of SU(2)}

\author{Kevin Cahill\address{Department of Physics and Astronomy,
        University of New Mexico, \\
        Albuquerque, New Mexico 87131-1156, U.~S.~A.}%
                \thanks{Research supported
                by the U.~S. Department of Energy
                under contract DE-FG04-84ER40166;
                e-mail: kevin@cahill.unm.edu.}}

\begin{document}

\begin{abstract}
Wilson loops have been measured
at strong coupling, $\beta=0.5$, on a $12^4$ lattice
in a noncompact simulation of pure SU(2)
in which random compact gauge transformations impose
a kind of lattice gauge invariance.
The Wilson loops
suggest a confining potential.
\end{abstract}

\maketitle

\section{INTRODUCTION}

In 1980 Creutz~\cite{Creu80a}
displayed quark
confinement at moderate coupling
in lattice simulations of both abelian and nonabelian gauge theories.
Whether nonabelian confinement
is as much an artifact of Wilson's action
as is abelian confinement remains unclear.
The basic variables of
Wilson's formulation~\cite{Wils74}
are elements of a compact group
and enter the action only through
traces of their products.
The Wilson action has false vacua~\cite{Cahi88}
which affect the string tension~\cite{Mack82,Grad88}.
\par
In simulations of SU(2)
with gauge-invariant
potential barriers
between the true vacuum and the false vacua,
the string tension has been seen to drop~\cite{Mack82}
or even vanish\cite{Grad88}.
\par
To examine these questions,
some physicists have introduced lattice actions
that are noncompact discretizations
of the continuum action
with fields as the basic variables.
For U(1) these noncompact formulations are
accurate for all coupling strengths~\cite{Cahi86};
for SU(2) they agree well with perturbation theory
at very weak coupling~\cite{Cahi89}.
\par
This report relates the results of measuring
Wilson loops at strong coupling, $\beta\equiv 4/g^2=0.5$,
on a $12^4$ lattice
in a noncompact simulation of SU(2) gauge theory
without gauge fixing or fermions.
In this simulation the fields are subjected
to random compact gauge transformations
which restore a semblance of
lattice gauge invariance.
\bigskip
\par
\section{NONCOMPACT METHODS}
Patrascioiu {\it et al.\/} performed
the first noncompact simulations of SU(2)
by discretizing the classical action
and fixing the gauge~\cite{Patr81}.
They saw a Coulomb force.
\par
Later simulations~\cite{Cahi93}
were carried out with
an action free of spurious zero modes,
for which it was not necessary to fix the gauge.
The Wilson loops of these simulations showed
no sign of quark confinement.
A possible explanation of this negative result
is that noncompact actions
lack an exact lattice gauge invariance.
Yet if one subjects
the fields to random compact
gauge transformations during each sweep,
then one may be able to restore a kind of
gauge invariance to the simulation~\cite{Cahi93}.
Here I report the results of such a gauge-invariant
simulation in which the Wilson loops fall off
exponentially with the area of the loop.
\par
In both the earlier simulations without
gauge invariance
and the new simulation
with gauge invariance,
the fields are constant on the links
of length $a$, the lattice spacing, but are interpolated
linearly throughout the plaquettes.
In the plaquette with vertices $n$, $n+e_\mu$,
$n+e_\nu$, and $n+e_\mu+e_\nu$,
the field is
\begin{eqnarray}
A_\mu^a(x) & = & ({x_\nu \over a} -n_\nu) A_\mu^a(n+e_\nu) \nonumber \\
& & \> + \, (n_\nu +1-{x_\nu \over a})A_\mu^a(n),
\end{eqnarray}
and the field strength is
\begin{eqnarray}
F_{\mu\nu}^a(x) & = & \partial_\nu A_\mu^a(x)-\partial_\mu A_\nu^a(x)
\nonumber \\
& & \> + \, g f_{bc}^a A_\mu^b(x) A_\nu^c(x).
\end{eqnarray}
The action $S$ is the sum over all plaquettes
of the integral over each plaquette of the
squared field strength,
\begin{equation}
S=\sum_{p_{\mu\nu}}
{a^2 \over 2} \int \! dx_\mu dx_\nu F_{\mu\nu}^c(x)^2.
\end{equation}
The mean value in the vacuum of
a euclidean-time-ordered operator $Q$
is approximated by a ratio of multiple integrals
over the $A_\mu^a(n)$'s
\begin{equation}
\langle {\cal T} Q(A) \rangle_0  \approx
{\int
e ^ {-S(A)}  Q(A)
\prod_{\mu,a,n}  dA_\mu^a(n)
\over \strut
 \int
e ^ {-S(A)}
\prod_{\mu,a,n}  dA_\mu^a(n)
}
\end{equation}
which one may compute numerically.
Macsyma was used to write most of
the Fortran code~\cite{Cahi90}
for the present simulation.
\par
\section{GAUGE INVARIANCE}
To restore gauge invariance,
the fields are subjected to
random compact gauge transformations
during every sweep, except those devoted
exclusively to measurements.
At each vertex $n$ a random
number $r$ is generated uniformly on the interval $(0,1)$;
and if $r$ is less than a fixed probability,
set equal to 0.5 in this work,
then a random group element $U(n)$
is picked from the group $SU(2)$.
The fields on the four links coming
out of the vertex $n$ are then subjected to the
compact gauge transformation
\begin{equation}
e^{-igaA_\mu^{\prime b}(n)T_b} =
e^{-igaA_\mu^a(n)T_a}U(n)^\dagger
\end{equation}
and those on the links entering the vertex
to the transformation
\begin{equation}
e^{-igaA_\mu^{\prime b}(n-e_\mu)T_b} =
U(n) e^{-igaA_\mu^a(n-e_\mu)T_a}.
\end{equation}
\section{WILSON LOOPS}
The quantity normally used to study confinement
in quarkless gauge theories is the Wilson loop $W(r,t)$
which is the mean value in the
vacuum of the trace of a path-and-time-ordered
exponential of a line integral of the connection
around an $r\times t$ rectangle
\begin{equation}
W(r,t) = (1/d) \, \langle {\rm tr \> }
{\cal PT} e^{-i g\oint \! A_\mu^a T_a dx_\mu}
\rangle_0
\end{equation}
where $d$ is the dimension of the generators $T_a$.
Although Wilson loops vanish
in the exact theory~\cite{Cahi79},
Creutz ratios $\chi (r,t)$
of Wilson loops defined~\cite{Creu80b}
as double differences of logarithms of Wilson loops
are finite. For large $t$, $\chi (r,t)$ approximates
($a^2$ times) the force between a quark
and an antiquark separated by the distance $r$.
\par
In this simulation
the data are not yet sufficient to allow one to
determine the Creutz ratios beyond the $3\times4$ loop.
The Wilson loops therefore have been
fitted to an expression involving Coulomb, perimeter,
scale, and area terms.
\section{MEASUREMENTS AND RESULTS}
It will be useful to compare this simulation
with an earlier one~\cite{Cahi93}
in which the fields were not subjected
to random gauge transformations.
Both simulations were done on
a $12^4$ periodic lattice with a heat bath.
The earlier simulation consisted of
20 independent runs with cold starts.
The first run had 25,000 thermalizing
sweeps at inverse coupling $\beta = 2$
followed by 5000 at $\beta=0.5$;
the other nineteen runs began at $\beta = 0.5$
with 20,000 thermalizing sweeps.
There were 59,640 Parisi-assisted~\cite{Pari83}
measurements, 20 sweeps apart.
\par
$$
\vbox{
\hbox{\it Noncompact Wilson loops at $\beta = 0.5$}
\vskip 1pt
\vbox{\offinterlineskip
\hrule
\halign{&\vrule#&
  \strut\quad#\hfil\quad\cr
height2pt&\omit&&\omit&&\omit&\cr
&${r\over a}\times{t\over a}$\hfil&&Not invariant\hfil&&
Invariant\hfil&\cr
height2pt&\omit&&\omit&&\omit&\cr \noalign{\hrule}
height2pt&\omit&&\omit&&\omit&\cr
&$1\times1$&&0.402330(6)&&0.254564(8)&\cr
height2pt&\omit&&\omit&&\omit&\cr \noalign{\hrule}
height2pt&\omit&&\omit&&\omit&\cr
&$2\times2$&&0.085426(4)&&0.018711(6)&\cr
height2pt&\omit&&\omit&&\omit&\cr \noalign{\hrule}
height2pt&\omit&&\omit&&\omit&\cr
&$3\times3$&&0.018080(2)&&0.001429(3)&\cr
height2pt&\omit&&\omit&&\omit&\cr \noalign{\hrule}
height2pt&\omit&&\omit&&\omit&\cr
&$4\times4$&&0.003993(1)&&0.000117(3)&\cr
height2pt&\omit&&\omit&&\omit&\cr \noalign{\hrule}
height2pt&\omit&&\omit&&\omit&\cr
&$5\times5$&&0.000893(1)&&0.000014(6)&\cr
height2pt&\omit&&\omit&&\omit&\cr \noalign{\hrule}
height2pt&\omit&&\omit&&\omit&\cr
&$6\times6$&&0.000201(0)&&0.000004(2)&\cr
height2pt&\omit&&\omit&&\omit&\cr}
\hrule}}
$$
\par
The present simulation with random gauge transformations
is very noisy.
So far it consists of 13 runs, all with cold starts
and 20,000 thermalizing sweeps.
Wilson loops have been measured every five sweeps
for a total of 689,684 measurements.
The values of the diagonal Wilson loops so obtained
are listed in the table.
The errors have been estimated
by the jackknife method,
with all measurements in bins of 100
considered to be independent.
\begin{figure} [htb]
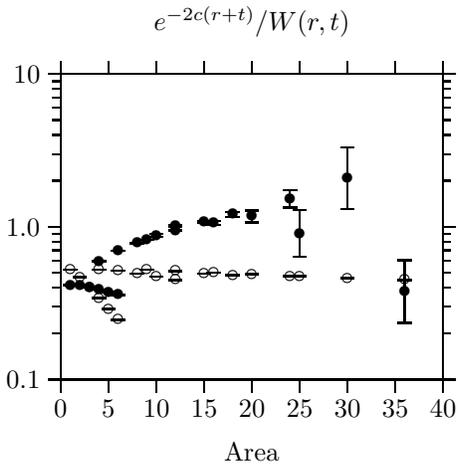

\beginpicture
\inboundscheckon
\setcoordinatesystem units <0.05in,0.8in>
\setplotarea x from 0 to 40, y from -1 to 1
\axis bottom label {Area} ticks
  numbered from 0 to 40 by 5 /
\axis left label {}
  ticks logged
  numbered at 0.1 1.0 10 /
  unlabeled short from 0.2 to 0.9 by 0.1
  from 2 to 9 by 1 /
\axis top label {$e^{-2c(r+t)}/W(r,t)$}
  ticks
  unlabeled from 0 to 40 by 5 /
\axis right ticks logged
unlabeled at 0.1 1.0 10 /
   short from 0.2 to 0.9 by 0.1
  from 2 to 9 by 1 /

\inboundscheckon

\crossbarlength=5pt

\puterrorbar at   1 -0.382476217 with fuzz  0.000013802
\puterrorbar at   2 -0.381644841 with fuzz  0.000043726
\puterrorbar at   4 -0.225461964 with fuzz  0.000144576
\puterrorbar at   3 -0.396154434 with fuzz  0.000136126
\puterrorbar at   6 -0.159470859 with fuzz  0.000120592
\puterrorbar at   9 -0.084928613 with fuzz  0.001053651
\puterrorbar at   4 -0.412730727 with fuzz  0.000275677
\puterrorbar at   8 -0.107325745 with fuzz  0.002533451
\puterrorbar at  12 -0.024758703 with fuzz  0.001523808
\puterrorbar at  16  0.024143035 with fuzz  0.012665630
\puterrorbar at   5 -0.428910380 with fuzz  0.000818503
\puterrorbar at  10 -0.058234294 with fuzz  0.008680739
\puterrorbar at  15  0.031234096 with fuzz  0.005238929
\puterrorbar at  20  0.066619707 with fuzz  0.039169468
\puterrorbar at  25 -0.044406170 with fuzz  0.152605727
\puterrorbar at   6 -0.447276890 with fuzz  0.003573211
\puterrorbar at  12  0.004423093 with fuzz  0.004436861
\puterrorbar at  18  0.082250082 with fuzz  0.015743524
\puterrorbar at  24  0.183570772 with fuzz  0.056907520
\puterrorbar at  30  0.316508633 with fuzz  0.201463744
\puterrorbar at  36 -0.424861664 with fuzz  0.205404639

\putcirclebar at   1 -0.279137910 with fuzz  0.000006153
\putcirclebar at   2 -0.330097160 with fuzz  0.000010560
\putcirclebar at   4 -0.280703147 with fuzz  0.000022826
\putcirclebar at   3 -0.397522318 with fuzz  0.000019929
\putcirclebar at   6 -0.288538310 with fuzz  0.000032347
\putcirclebar at   9 -0.280841744 with fuzz  0.000057889
\putcirclebar at   4 -0.467791075 with fuzz  0.000035489
\putcirclebar at   8 -0.307249212 with fuzz  0.000047242
\putcirclebar at  12 -0.291248054 with fuzz  0.000072399
\putcirclebar at  16 -0.299547960 with fuzz  0.000135925
\putcirclebar at   5 -0.538561918 with fuzz  0.000062234
\putcirclebar at  10 -0.327958623 with fuzz  0.000069811
\putcirclebar at  15 -0.304707734 with fuzz  0.000104236
\putcirclebar at  20 -0.311795485 with fuzz  0.000163179
\putcirclebar at  25 -0.323634392 with fuzz  0.000330575
\putcirclebar at   6 -0.609341776 with fuzz  0.000109557
\putcirclebar at  12 -0.348953475 with fuzz  0.000104815
\putcirclebar at  18 -0.318497862 with fuzz  0.000158418
\putcirclebar at  24 -0.324971186 with fuzz  0.000242348
\putcirclebar at  30 -0.336282222 with fuzz  0.000390073
\putcirclebar at  36 -0.349576837 with fuzz  0.000843690

\endpicture
\caption{The negative logarithms
of Wilson loops with the
perimeter factor canceled
are plotted against the area $rt$ of the loop.
The loops of the gauge-invariant simulation
are represented by bullets; those of the
earlier simulation by circles.}
\end{figure}
\par
The Wilson loops of the gauge-invariant
simulation fall off much faster with increasing loop size
than do those of the earlier simulation.
Because the data do not accurately determine
all the Creutz ratios, I have fitted both
sets of loops, including the non-diagonal loops,
to the formula
\begin{equation}
W(r,t) \approx e^{ a + b(r/t + t/r) - 2c\,(r + t) - d\, rt}
\end{equation}
in which $a$ is a scale factor,
$b$ a Coulomb term, $c$ a perimeter term,
and $d$ an area term.
For the simulation without random gauge transformations,
I found $a\approx 0.25$, $b\approx 0.20$,
$c\approx 0.39$, and $d\approx 0.00$.
For the simulation with random gauge transformations,
I found $a\approx 0.60$, $b\approx 0.20$,
$c\approx 0.56$, and $d\approx 0.11$.
In the gauge-invariant simulation,
the coefficient of the area-law term is
about two orders of magnitude larger than in the earlier
simulation which lacked gauge invariance.
\par
To exhibit the renormalized quark-antiquark potential,
I have plotted in the figure the negative logarithms
$-\log_{10} \left(e^{2c(r+t)}W(r,t)\right)$ of the Wilson
loops with the perimeter term removed.
Apart from the uncertain value of $W(6,6)$,
the bigger loops of the gauge-invariant simulation,
represented by bullets, display an area law;
whereas the larger loops of the earlier simulation,
represented by circles, show an essentially flat potential.
The smallest loops reflect the symmetrized Coulomb term
$\propto (t/r + r/t)$.
\par
\centerline{Acknowledgments}
I should like to thank
H.~Barnum for participating
in some of the early stages of this work;
M.~Creutz, G.~Herling, G.~Kilcup, J.~Polonyi, and D.~Topa
for useful conversations;
and the Department of Energy for support under grant
DE-FG04-84ER40166.
Computer time was generously supplied by
the UNM neutrino group,
by the UNM high-energy-physics group,
by IBM, by NERSC, and by Richard Matzner's
Cray Research grant at the
Center for High-Performance Computing
of the Univ.\ of Texas System.


\begin{thebibliography}{9}
\bibitem{Creu80a} M.~Creutz, {\it Phys. Rev. D\/} 21 (1980) 2308;
                 {\it Phys.~Rev.~Letters} 45 (1980) 313.
\bibitem{Wils74} K.~Wilson, {\it Phys.~Rev.~D\/} 10 (1974) 2445.
\bibitem{Cahi88} K.~Cahill, M.~Hebert, and S.~Prasad,
                 {\it Phys.\ Lett.\ B\/} 210 (1988) 198;
                 K.~Cahill and S.~Prasad,
                 {\it Phys.~Rev.~D\/} 40 (1989) 1274.
\bibitem{Mack82} G.~Mack and E.~Pietarinen,
                 {\it Nucl.~Phys.~B\/} 205 (1982) 141.
\bibitem{Grad88} M.~Grady, {\it Z.~Phys.~C\/} 39 (1988) 125.
\bibitem{Cahi86} K.~Cahill and R.~Reeder,
                 {\it Phys.~Lett.~B\/} 168 (1986) 381;
                 {\it J.~Stat.~Phys.\/} 43 (1986) 1043.
\bibitem{Cahi89} K.~Cahill,
                 {\it Nucl.~Phys.~B (Proc.~Suppl.)\/} 9 (1989) 529;
                 {\it Phys. Lett.~B\/} 231 (1989) 294.
\bibitem{Patr81} A.~Patrascioiu, E.~Seiler, and I.~Stamatescu,
                 {\it Phys.~Lett.~B\/} 107 (1981) 364;
                 I.~Stamatescu, U.~Wolff and D.~Zwanziger,
                 {\it Nucl.~Phys.~B\/} 225 [FS9] (1983) 377;
                 E.~Seiler, I.~Stamatescu, and D.~Zwanziger,
                 {\it Nucl.~Phys.~B\/} 239 (1984) 177 and 201.
\bibitem{Cahi93} K.~Cahill,
                 {\it Phys.~Lett.\/} B304 (1993) 307.
\bibitem{Cahi90} K.~Cahill,
                 {\it Comput.~Phys.\/} 4 (1990) 159.
\bibitem{Cahi79} K.~Cahill and D.~Stump,
                 {\it Phys.~Rev.~D\/} 20 (1979) 2096.
\bibitem{Creu80b} M.~Creutz,
                 {\it Phys. Rev. D\/} { 21} (1980) 2308;
                 {\it Phys.~Rev.~Letters} 45 (1980) 313.
\bibitem{Pari83} G.~Parisi, R.~Petronzio, and F.~Rapuano,
                 {\it Phys.\ Lett.\ B\/} 128 (1983) 418.
\end{thebibliography}
\end{document}